\documentclass[prb,twocolumn,preprintnumbers,amsmath,amssymb,floatfix,showpacs]{revtex4}

\usepackage[dvips]{graphics,graphicx,color}

\begin{document}

\title{Quantum Monte Carlo calculation of the energy band and
quasiparticle effective mass of the two-dimensional Fermi fluid}

\author{N.\ D.\ Drummond and R.\ J.\ Needs}

\affiliation{TCM Group, Cavendish Laboratory, University of Cambridge,
J.\ J.\ Thomson Avenue, Cambridge CB3 0HE, United Kingdom}

\begin{abstract} We have used the diffusion quantum Monte Carlo method
to calculate the energy band of the two-dimensional homogeneous
electron gas (HEG), and hence we have obtained the quasiparticle
effective mass and the occupied bandwidth. We find that the effective
mass in the paramagnetic HEG increases significantly when the density
is lowered, whereas it decreases in the fully ferromagnetic HEG\@.
Our calculations therefore support the conclusions of recent
experimental studies [Y.-W.\ Tan \textit{et al.}, Phys.\ Rev.\ Lett.\
\textbf{94}, 016405 (2005); M.\ Padmanabhan \textit{et al.}, Phys.\
Rev.\ Lett.\ \textbf{101}, 026402 (2008); T.\ Gokmen \textit{et al.},
Phys.\ Rev.\ B \textbf{79}, 195311 (2009)]. We compare our calculated
effective masses with other theoretical results and experimental
measurements in the literature.
\end{abstract}


\maketitle

\section{Introduction \label{sec:introduction}}

Landau's Fermi liquid theory\cite{landau} is an immensely successful
and widely used framework for understanding the properties of
interacting electron systems.\cite{giuliani} Low-energy excitations in
a fluid of interacting electrons can be treated as excitations of
independent quasiparticles, whose energy-momentum relationship (the
quasiparticle energy band) generally differs from that of free
electrons.  Close to the Fermi surface, the quasiparticle band can be
approximated by the free-particle form appropriate for particles of
mass $m^\ast$, where the \textit{quasiparticle effective mass}
$m^\ast$ may differ from the bare mass of an electron.  Given the
widespread use of Fermi liquid theory, it comes as something of a
surprise to learn that the effective mass of a paramagnetic
two-dimensional (2D) homogeneous electron gas (HEG) has been the
subject of great controversy in recent years.  Early
experiments\cite{smith,pudalov} found a large enhancement of the
effective mass at low density, but subsequent
experiments\cite{tan2005,padmanabhan} have found the increase in the
effective mass to be considerably smaller.  On the theoretical side,
many-body perturbation theory ($GW$) calculations give a range of
possible results depending on the choice of effective interaction and
whether or not the Dyson equation is solved
self-consistently,\cite{asgari2005,asgari_2006,giuliani} while quantum
Monte Carlo (QMC) studies have found either much
less\cite{kwon_fermi_liquid,kwon_trans_est} or much
greater\cite{holzmann} enhancement of the effective mass than the
experiments suggest.  Finally, some recent
experiments\cite{padmanabhan} have shown that the effective masses in
paramagnetic and ferromagnetic HEGs behave quite differently as a
function of density, as had been predicted using many-body
perturbation theory.\cite{das_sarma} The experiments show that the
effective mass of a ferromagnetic HEG decreases as the density is
lowered, which has also been observed in a recent $GW$
study.\cite{asgari_2009} Understanding the magnetic behavior of the 2D
HEG at low density will play an important role in the design of
spintronic devices.

In this article, we present QMC calculations\cite{foulkes_2001} of the
energy band of the 2D HEG\@.  We have calculated the band ${\cal
E}(k)$ by evaluating the difference in the total energy when an
electron is added to or removed from a particular momentum state ${\bf
k}$.  To our knowledge, this is the first QMC calculation of the
complete 2D HEG occupied energy band.  As explained above, electronic
excitations close to the Fermi surface correspond to quasiparticle
excitations. The electronic and quasiparticle bands therefore agree
near the Fermi surface and have the same derivative at $k_F$.  The
effective mass of a HEG can be written as\cite{giuliani}
$m^\ast=k_F/(\partial {\cal E} / \partial k)_{k_F}$, where $k_F$ is
the Fermi wave vector,\cite{footnote_k_F} and hence it is
straightforward to compute the effective mass once the energy band has
been determined.  Our effective-mass data should help to resolve the
controversies surrounding the increase in the effective mass of the
paramagnetic 2D HEG at low density.  We have studied both paramagnetic
and ferromagnetic HEGs in order to look for the differences in
behavior observed by Padmanabhan \textit{et al.}\cite{padmanabhan}

We use Hartree atomic units ($\hbar=|e|=m_e=4\pi\epsilon_0=1$)
throughout.  Densities are given in terms of the radius $r_s$ of the
circle that contains one electron on average.  All our QMC
calculations were performed using the \textsc{casino}
code.\cite{casino}

The rest of this article is arranged as follows.  We explain our
methodology in Sec.\ \ref{sec:methodology}.  We present our results in
Sec.\ \ref{sec:results}. Finally we draw our conclusions in Sec.\
\ref{sec:conclusions}.

\section{Methodology \label{sec:methodology}}

In the variational quantum Monte Carlo (VMC) method, expectation
values are calculated with respect to a trial wave function, the
integrals being performed by a Monte Carlo technique.  In the
diffusion quantum Monte Carlo\cite{foulkes_2001} (DMC) method the
imaginary-time Schr\"odinger equation is used to evolve an ensemble of
electronic configurations towards the ground state.  Fermionic
symmetry is maintained by the fixed-node
approximation,\cite{anderson_1976} in which the nodal surface of the
wave function is constrained to equal that of a trial wave function.

Our trial wave functions consisted of Slater determinants of
plane-wave orbitals multiplied by a Jastrow correlation factor. The
Jastrow factor contained polynomial and plane-wave expansions in
electron-electron separation.\cite{ndd_jastrow} The orbitals in the
Slater wave function were evaluated at quasiparticle coordinates
related to the actual electron positions by backflow functions
consisting of polynomial expansions in electron-electron
separation.\cite{backflow} The wave functions were optimized by
variance minimization\cite{umrigar_1988a,ndd_newopt} followed by
linear-least-squares energy minimization.\cite{umrigar_emin}  The high
quality of our trial wave functions is demonstrated in this paper and
in Ref.\ \onlinecite{ndd_2dheg_expvals}.

The single-particle energy for an occupied state at wave vector ${\bf
k}$ is defined to be ${\cal E}({\bf k}) \equiv E_0-E_-({\bf k})$,
while the single-particle energy for an unoccupied state is ${\cal
E}({\bf k}) \equiv E_+({\bf k})-E_0$, where $E_0$ is the ground-state
total energy, $E_+({\bf k})$ is the total energy of the system with an
extra electron placed in orbital $\exp(i{\bf k}\cdot {\bf
r})$,\cite{footnote_excitation} and $E_-({\bf k})$ is the total energy
with an electron removed from orbital $\exp(i{\bf k} \cdot {\bf r})$.
In a finite simulation cell subject to periodic boundary conditions,
the available states $\{{\bf k}\}$ fall on the grid of
reciprocal-lattice points offset by the simulation-cell Bloch vector
${\bf k}_s$.\cite{rajagopal_1994,rajagopal_1995} The simulation cell
was left unchanged when electrons were added or removed.  We have
confirmed that finite-size biases are negligible by carrying out
simulations in different cell sizes: see Figs.\ \ref{fig:para_bands}
and \ref{fig:ferro_bands}.  We have also verified that the band values
obtained with different simulation-cell Bloch vectors lie on the same
curve as the band values obtained with ${\bf k}_s={\bf 0}$.  Having
determined the energy band at a series of $k$ values, we performed a
least-squares fit of a quartic function ${\cal E}(k)=\alpha_0+\alpha_2
k^2+\alpha_4 k^4$ to the band values.  The DMC energy band is defined
as a difference in total-energy eigenstates; as explained in the
introduction, this coincides with the quasiparticle band near the
Fermi surface and hence gives a correct description of the effective
mass.

The number of electrons $N$ in each of our ground-state calculations
was chosen to be a ``magic number'' corresponding to a closed-shell
configuration when ${\bf k}_s={\bf 0}$.  In this case real,
single-determinant wave functions are appropriate for the ground-state
calculations, facilitating the optimization of the wave function.  In
the $(N+1)$- and $(N-1)$-electron excited-state calculations, we used
the Jastrow factor and backflow function that were optimized for the
$N$-electron ground state.  Reoptimizing the wave function in the
excited state was not found to make a significant difference to the
VMC or DMC energies. To try to obtain a better estimate of the energy
of the $(N+1)$-electron system, we constructed a multideterminant wave
function in which the extra electron occupied each of the
symmetry-equivalent ${\bf k}$ vectors in the partially filled
shell. The determinant coefficients were free parameters, which we
optimized by linear-least-squares energy minimization.  However, we
were unable to lower the VMC energy significantly using this form of
wave function, so in our production calculations we used
single-determinant wave functions for the $(N+1)$- and
$(N-1)$-electron systems.  Our DMC results are converged with respect
to time step, as is clear from the agreement between the energy bands
obtained at different time steps in Fig.\ \ref{fig:para_bands}(b).

Unlike the DMC calculations of Kwon \textit{et
al.},\cite{kwon_fermi_liquid,kwon_trans_est} we did not promote
electrons from the ground-state configuration; we simply added or
subtracted single electrons.  The energy difference that results from
promoting an electron contains a contribution from the interaction
between the excited electron and the hole that it leaves behind, in
addition to the difference of band energies.  By contrast, the energy
difference resulting from adding or subtracting an electron simply
gives the corresponding band energy.  We believe our approach to be a
simpler procedure for calculating the band and hence the effective
mass, although it does not give values for the
quasiparticle-interaction Fermi-liquid parameters.  Our determination
of the effective mass will facilitate subsequent calculation of the
other Fermi-liquid parameters using the approach of Kwon \textit{et
al.}

The other important difference between our methodology and that of
Kwon \textit{et al.}\ is that we have used a fit to the entire
occupied band to determine the derivative at $k_F$ and hence the
effective mass.  This was done for the following reasons: (i)
evaluating the derivative numerically using only a few band values
near $k_F$ is unreliable because of the noise in the band data; (ii)
the DMC-calculated band suffers from Hartree-Fock-like pathological
behavior in the vicinity of $k_F$ because the method does not retrieve
all the correlation energy (see the discussion in Sec.\
\ref{sec:em_results}); and (iii) the band evaluated in a finite cell
may suffer from finite-size effects in the vicinity of the Fermi
surface.\cite{holzmann} Although the pathological behavior dominates
the derivative of the band in the vicinity of $k_F$, it has only a
negligible effect on the band fitted over a wide range of $k$.  The
fit to the band is good, so that the derivative of the fitted band at
$k_F$ should be reliable: see Figs.\ \ref{fig:para_bands} and
\ref{fig:ferro_bands} and Figs.\ \ref{fig:para_dbands} and
\ref{fig:ferro_dbands}.

The occupied bandwidth of the HEG is $\Delta {\cal E}={\cal
E}({k_F})-{\cal E}(0)=E_-(0)-E_-(k_F)$.  The DMC bandwidth is expected
to be an upper bound on the true bandwidth: assuming that DMC
retrieves the same fraction of the correlation energy in the ground
and excited states, the bandwidth will lie between the Hartree-Fock
(HF) value $E_-^{\rm HF}(0)-E_-^{\rm HF}(k_F)$, which is too
large,\cite{giuliani} and the exact result $E_-^{\rm
exact}(0)-E_-^{\rm exact}(k_F)$. Likewise, the Slater-Jastrow DMC
bandwidths are expected to be greater than the Slater-Jastrow-backflow
DMC bandwidths, as can be seen to be the case in Fig.\
\ref{fig:para_bands}(b). To obtain an accurate bandwidth, it is
essential to retrieve a very large fraction of the correlation energy
in the QMC calculations, which explains why the use of DMC and the
inclusion of backflow is so important in this work.

\begin{figure}
\begin{center}
\includegraphics[scale=0.3,clip]{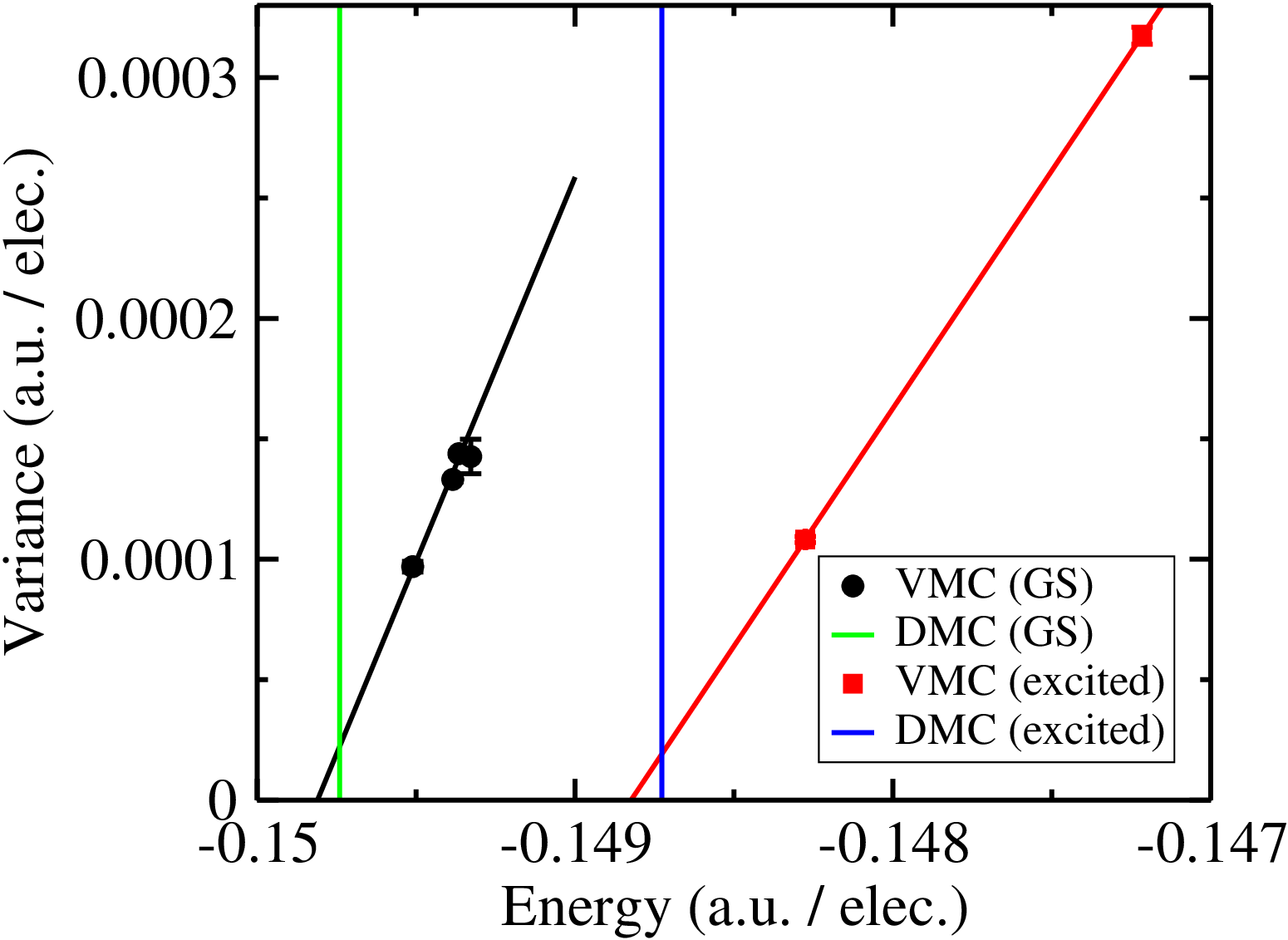}
\caption{(Color online) VMC variance against energy for different
trial wave functions for a 58-electron paramagnetic HEG of density
parameter $r_s=5$ a.u.  Plots are shown for the ground state (GS) and
an excited state in which an electron is removed from the
highest-occupied shell.  The slanted lines show fits to the VMC data.
The vertical lines show the fixed-node DMC energies obtained with
Slater-Jastrow-backflow wave functions.  The HF ground-state and
excited-state energies are $-0.100222$ and $-0.099632$ a.u.\ per
electron, respectively.
The difference between the slanted and vertical lines at zero variance
gives an approximation to the correlation energy missing in the DMC
calculation.  This is clearly small compared with the difference
between the HF and DMC energies.
\label{fig:E_var_extrap}}
\end{center}
\end{figure}

A crude way of estimating the ground-state energy is to plot the VMC
energy against the variance obtained with different trial wave
functions and extrapolate the VMC energy linearly to zero variance, as
shown in Fig.\ \ref{fig:E_var_extrap}.  This procedure suggests that
our DMC calculations retrieve more than 99\% of the correlation
energy, and that the fraction retrieved is similar in both the
ground-state and excited-state calculations. Suppose the free-electron
bandwidth is greater than or approximately equal to the exact
bandwidth (this is true for the ferromagnetic HEG and approximately
true for the paramagnetic HEG), so that the error in the HF bandwidth
is less than or approximately equal to $\Delta {\cal E}^{\rm
HF}-\Delta {\cal E}^{\rm free}=k_F(1-2/\pi)$.  Hence the error in the
DMC bandwidth is less than $0.01k_F(1-2/\pi)\approx 0.007/r_s$ for a
ferromagnetic HEG and less than about $0.01k_F(1-2/\pi)\approx
0.005/r_s$ for a paramagnetic HEG\@.  Since the bandwidth falls off as
$r_s^{-2}$, the error is more significant at large $r_s$.  In the
worst case, the paramagnetic HEG at $r_s=10$ a.u., this argument
suggests that DMC overestimates the bandwidth by about $9$\%.  The
errors in the other results are much smaller.  It seems reasonable to
assume that if the bandwidth is overestimated by a given amount then
the effective mass will be underestimated by a similar fraction.

\section{Results \label{sec:results}}

\subsection{Energy bands \label{sec:band_results}}

\begin{figure}
\begin{center}
\includegraphics[scale=0.81,clip]{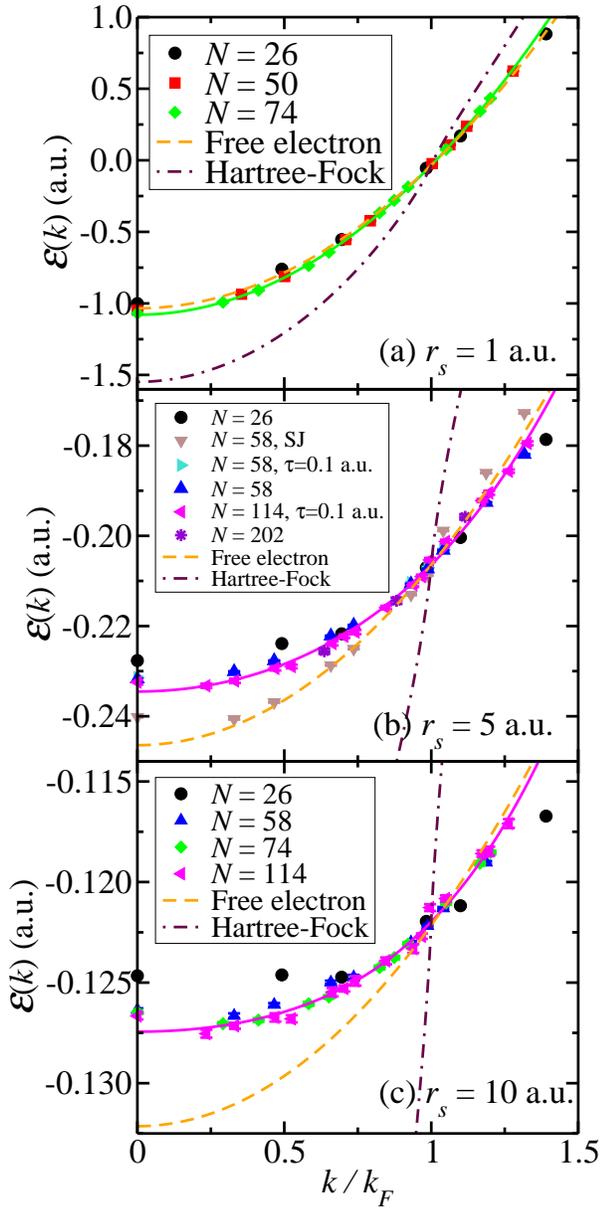}
\caption{(Color online) Energy bands of paramagnetic $N$-electron 2D
HEGs at $r_s=1$ a.u.\ (top), 5 a.u.\ (middle), and 10 a.u.\
(bottom). The free-electron and HF bands are offset to coincide with
the fitted DMC band at $k=k_F$. The curve labeled ``SJ'' used a
Slater-Jastrow trial wave function; the others used a
Slater-Jastrow-backflow trial wave function.  Except where indicated
otherwise, DMC time steps $\tau$ of 0.01, 0.2, and 0.4 a.u.\ were used
at $r_s=1$, 5, and 10 a.u.  The solid lines show quartic fits to the
DMC data for $N=74$, 114, and 114 electrons at $r_s=1$, 5, and 10
a.u., respectively.  (These are the largest system sizes for which we
have sufficient data to perform an adequate fit.)
\label{fig:para_bands}}
\end{center}
\end{figure}

\begin{figure}
\begin{center}
\includegraphics[scale=0.81,clip]{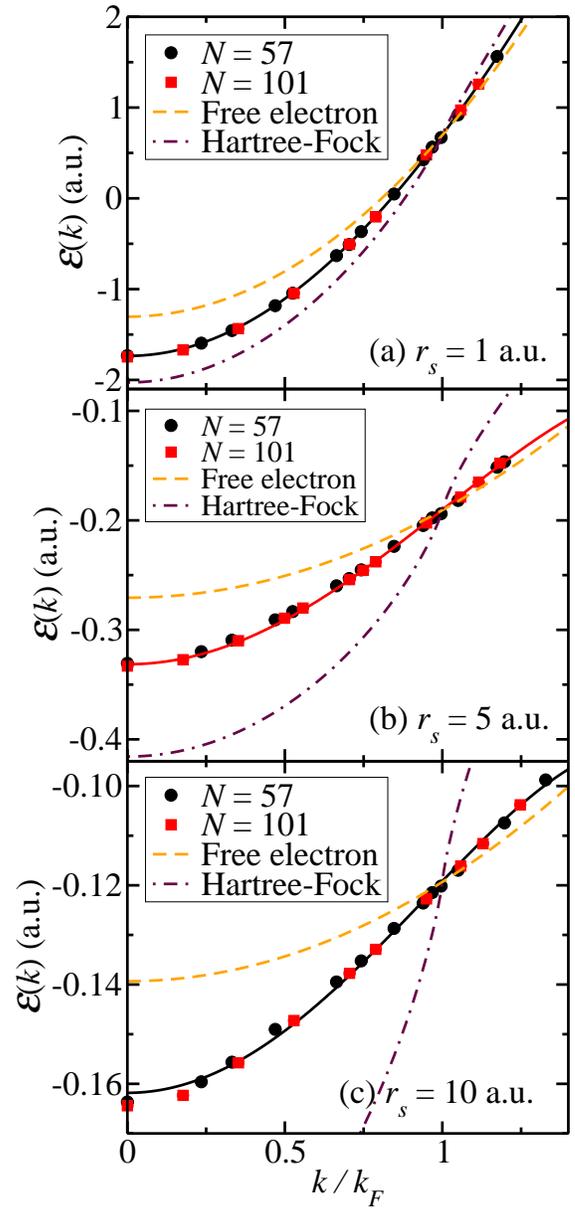}
\caption{(Color online) As Fig.\ \ref{fig:para_bands}, but for
ferromagnetic HEGs. The solid lines show quartic fits to the DMC data
for $N=57$, 101, and 57 electrons at $r_s=1$, 5, and 10 a.u.,
respectively.
  \label{fig:ferro_bands}}
\end{center}
\end{figure}

Our calculated energy bands are shown in Figs.\ \ref{fig:para_bands}
and \ref{fig:ferro_bands} for paramagnetic and ferromagnetic HEGs,
respectively.  The free-electron and HF bands are shown for
comparison.  As is well-known, the free-electron band is very much
more accurate than the HF band, especially at low densities and
especially in the paramagnetic fluid.  The HF band is pathological due
to the long range of the exchange hole, which results in incomplete
screening of the Coulomb interaction.\cite{giuliani}  Our DMC
bandwidths are shown in Table \ref{table:bw}.  The bandwidth in
paramagnetic HEGs at intermediate and low densities is less than the
free-electron bandwidth. On the other hand, in the ferromagnetic HEG
the bandwidth is greater than the free-electron bandwidth at all
densities studied.  In each case the DMC bandwidth is considerably
smaller than the HF bandwidth.

\begin{table}
\begin{center}
\begin{tabular}{cr@{.}lcr@{.}lcccccccc} \hline \hline

& \multicolumn{13}{c}{Bandwidth (a.u.)} \\

\raisebox{1ex}[0pt]{$r_s$} & \multicolumn{5}{c}{DMC} & &
\multicolumn{3}{c}{Free electron} & & \multicolumn{3}{c}{HF} \\

\raisebox{1ex}[0pt]{(a.u.)} & \multicolumn{2}{c}{Para.} & &
\multicolumn{2}{c}{Ferro.} & & Para. & & Ferro. & & Para. & &
Ferro. \\

\hline

~1 & $1$&$045(5)$ & & $2$&$434(6)$ & & $1.00$ & & $2.00$ & & $1.513$ &
& $2.726$ \\

~5 & $0$&$028\,1(8)$ & & $0$&$141(1)$ & & $0.04$ & & $0.08$ & &
$0.142$ & & $0.225$ \\

10 & $0$&$005\,5(3)$ & & $0$&$042\,7(8)$ & & $0.01$ & & $0.02$ & &
$0.061$ & & $0.092$ \\

\hline \hline
\end{tabular}
\caption{Bandwidths of paramagnetic and ferromagnetic 2D HEGs of
density parameter $r_s$, as calculated using DMC, free-electron theory
($\Delta {\cal E}=k_F^2/2$), and HF theory [$\Delta {\cal
E}=k_F^2/2+k_F(1-2/\pi)$].  The DMC bandwidths were obtained from the
fitted curves shown in Figs.\ \ref{fig:para_bands} and
\ref{fig:ferro_bands}.
\label{table:bw}}
\end{center}
\end{table}

It is striking how closely the DMC band agrees with the free-electron
band ${\cal E}(k)=k^2/2$ for a paramagnetic HEG at $r_s=1$ a.u.  At
$r_s=5$ and 10 a.u., the quartic term $\alpha_4k^4$ in the fitted band
is positive for the paramagnetic HEG\@.  For the ferromagnetic HEG the
quartic term is negative at all densities.  In either case the band is
clearly not quadratic.  This will result in non-free-particle-like
thermodynamic behavior at high temperatures.

\subsection{Effective masses \label{sec:em_results}}

\begin{figure}
\begin{center}
\includegraphics[scale=0.4,clip]{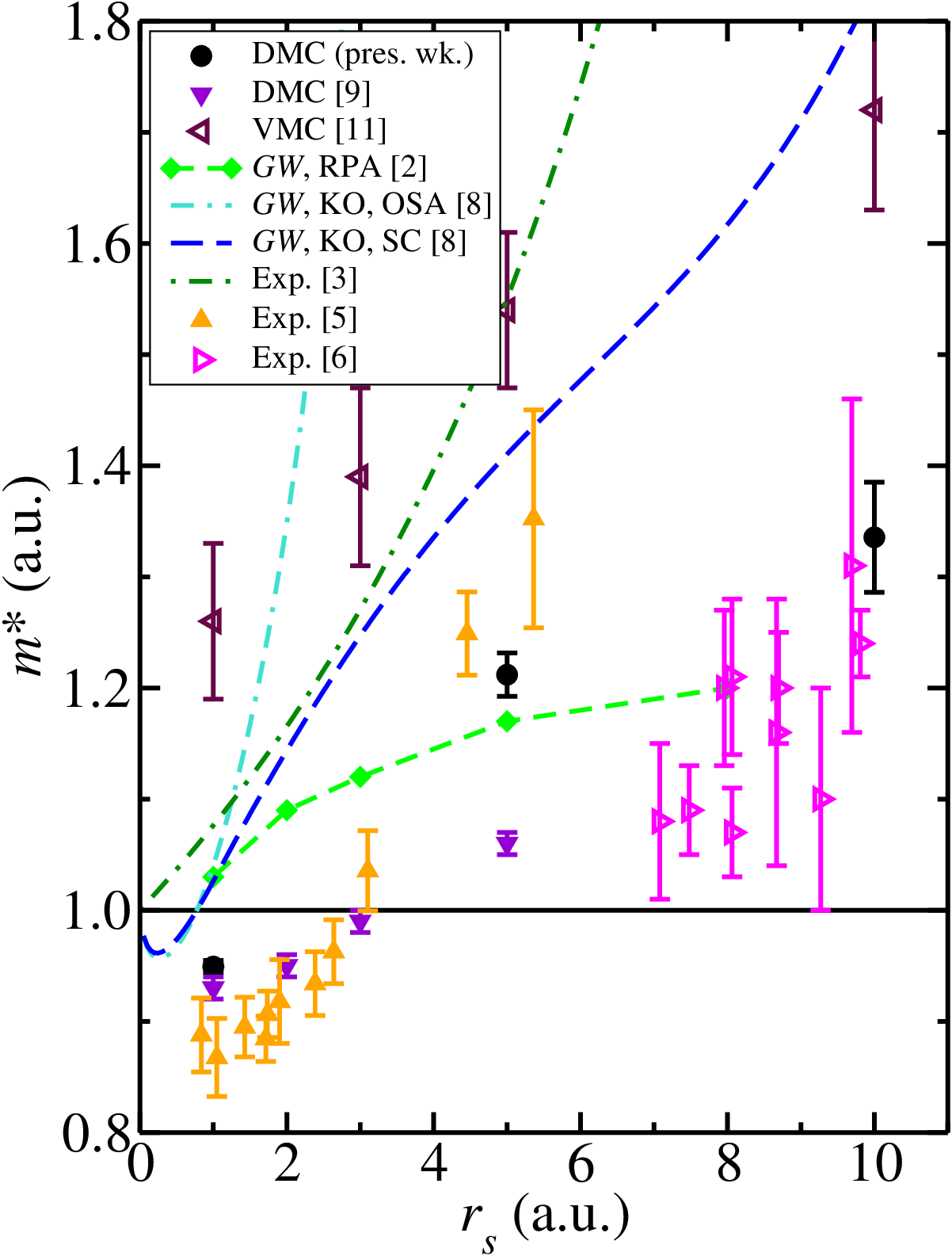}
\caption{(Color online) Effective mass $m^\ast$ against density
parameter $r_s$ for paramagnetic or partially spin-polarized 2D HEGs,
as calculated or measured by different authors. Our DMC results were
obtained from the fitted curves shown in Fig.\ \ref{fig:para_bands}.
The $GW$ results were obtained using the random-phase-approximation
(RPA) effective interaction\cite{giuliani} and the Kukkonen-Overhauser
(KO) effective interaction\cite{asgari_2006} by solving the Dyson
equation self-consistently (SC) or within the on-shell approximation
(OSA)\@. All the results shown are for paramagnetic HEGs with the
exception of the experimental results of Ref.\
\onlinecite{padmanabhan}, which are for a partially spin-polarized
HEG\@.  \label{fig:para_effmass}}
\end{center}
\end{figure}

\begin{figure}
\begin{center}
\includegraphics[scale=0.3,clip]{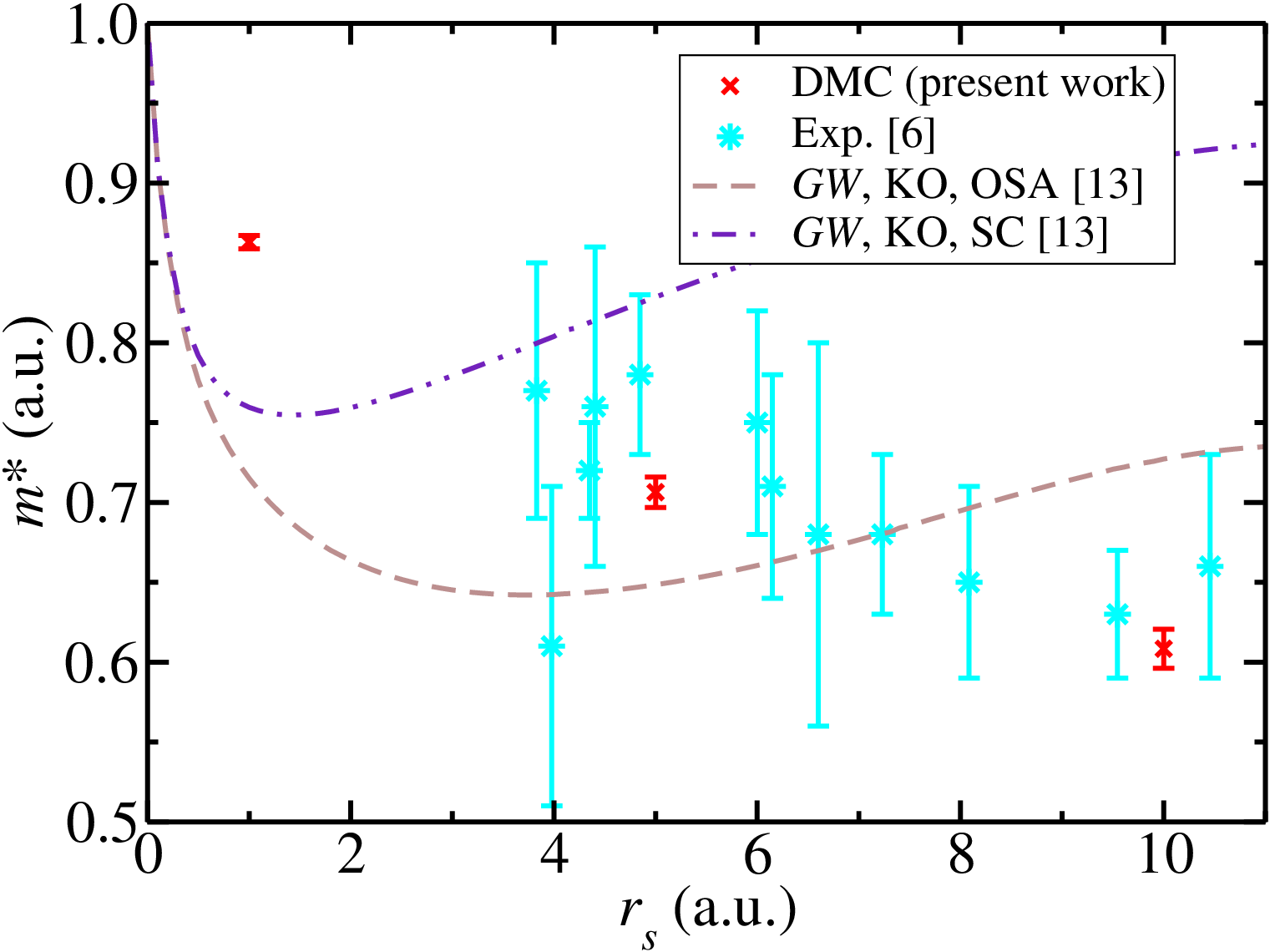}
\caption{(Color online) Effective mass $m^\ast$ against density
parameter $r_s$ for ferromagnetic 2D HEGs. Our DMC results were
obtained from the fitted curves shown in Fig.\
\ref{fig:ferro_bands}. The $GW$ results were obtained using the
Kukkonen-Overhauser (KO) effective interaction by solving the Dyson
equation self-consistently (SC) or within the on-shell approximation
(OSA)\@.\cite{asgari_2009}
\label{fig:ferro_effmass}}
\end{center}
\end{figure}

Our DMC effective masses for paramagnetic and ferromagnetic HEGs are
plotted in Figs.\ \ref{fig:para_effmass} and \ref{fig:ferro_effmass},
respectively, along with various experimental results and previous
theoretical predictions.  Our effective masses are also given in Table
\ref{table:eff_masses}.  In a paramagnetic HEG the effective mass
increases with $r_s$: at $r_s=1$ a.u.\ the effective mass is slightly
less than the bare electron mass, but at $r_s=5$ a.u.\ the effective
mass is significantly enhanced.  On the other hand, in ferromagnetic
HEGs the effective mass decreases when the density is lowered.  Our
results therefore support the conclusions of Padmanabhan \textit{et
al.}\cite{padmanabhan} In fact our ferromagnetic effective masses are
in good quantitative agreement with the experimental data of
Padmanabhan \textit{et al.},\cite{footnote_pad_mag} while our
paramagnetic effective masses are in reasonable agreement with those
measured by Tan \textit{et al.}\cite{tan2005} We do not find
especially good agreement with earlier theoretical work, however.  As
can be seen in Fig.\ \ref{fig:para_effmass}, the effective masses
obtained using the $GW$ method depend strongly on the choice of
effective interaction and whether or not the Dyson equation is solved
self-consistently, undermining confidence in that approach.  The DMC
data of Kwon \textit{et al.}\ do not show a significant enhancement of
the paramagnetic effective mass at low
densities.\cite{kwon_fermi_liquid} (Our calculations differ from those
of Kwon \textit{et al.}\ in that we use wave functions that retrieve a
greater fraction of the correlation energy, we use larger system
sizes, and we use different excitations to evaluate the effective
mass.)

\begin{table}
\begin{center}
\begin{tabular}{ccc} \hline \hline

& \multicolumn{2}{c}{$m^\ast$ (a.u.)}  \\

\raisebox{1ex}[0pt]{$r_s$ (a.u.)} & Para. & Ferro. \\

\hline

~1 & $0.949(6)$ & $0.863(4)$ \\

~5 & $1.21(2)~$ & $0.71(1)~$ \\

10 & $1.34(5)~$ & $0.61(1)~$ \\

\hline \hline
\end{tabular}
\caption{DMC-calculated effective masses $m^\ast$ of paramagnetic and
ferromagnetic 2D HEGs of density parameter $r_s$.
  \label{table:eff_masses}}
\end{center}
\end{table}

\begin{figure}
\begin{center}
\includegraphics[scale=0.3,clip]{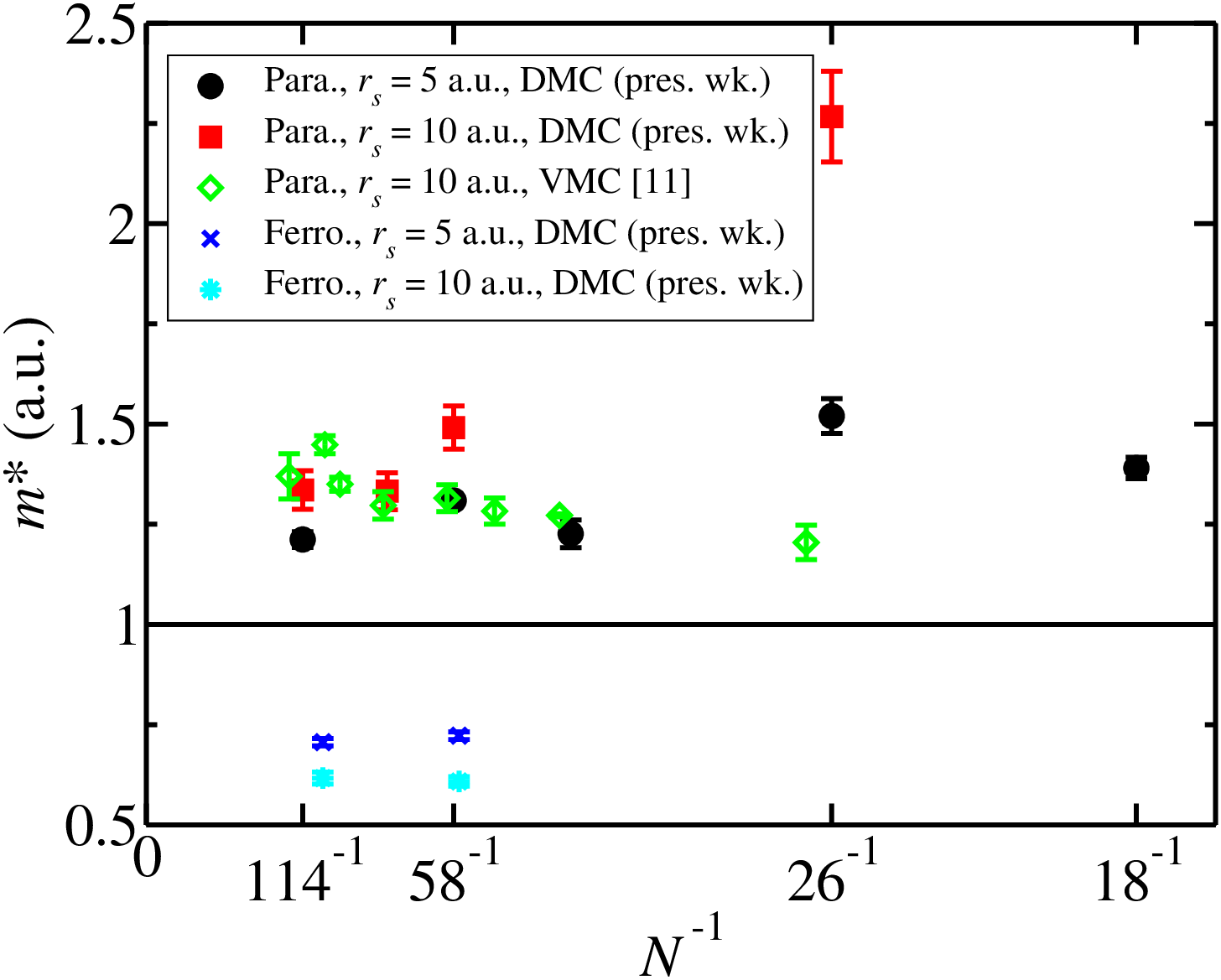}
\caption{(Color online) DMC effective mass $m^\ast$ against system
size $N$ for paramagnetic 2D HEGs.
\label{fig:para_qem_v_N}}
\end{center}
\end{figure}

Holzmann \textit{et al.}\cite{holzmann}\ have recently studied the
paramagnetic 2D HEG effective mass using the VMC method.  Their
effective masses differ significantly from our DMC results: see Fig.\
\ref{fig:para_effmass}.  Holzmann \textit{et al.}\ considered
additions of electrons in the vicinity of the Fermi surface.  They
found substantial finite-size effects in the effective mass, which
they corrected by considering the finite-size dependence of the
discontinuity in the momentum distribution at the Fermi edge.  Our
effective masses do not appear to suffer from these finite-size
effects; in fact, our effective masses show the \textit{opposite}
trend with system size, as can be seen in Fig.\
\ref{fig:para_qem_v_N}.  If the finite-size correction to the mass
falls off as $N^{-1/4}$ as predicted by Holzmann \textit{et al.}\ then
the correction should be roughly halved on going from $N=18$ to
$N=202$ electrons.  Noting that the correction is supposed to
\textit{increase} the mass, this ought to be visible in our
data. Since it is not, the finite-size correction of Holzmann
\textit{et al.}\ appears to be statistically insignificant.

\begin{figure}
\begin{center}
\includegraphics[scale=0.44,clip]{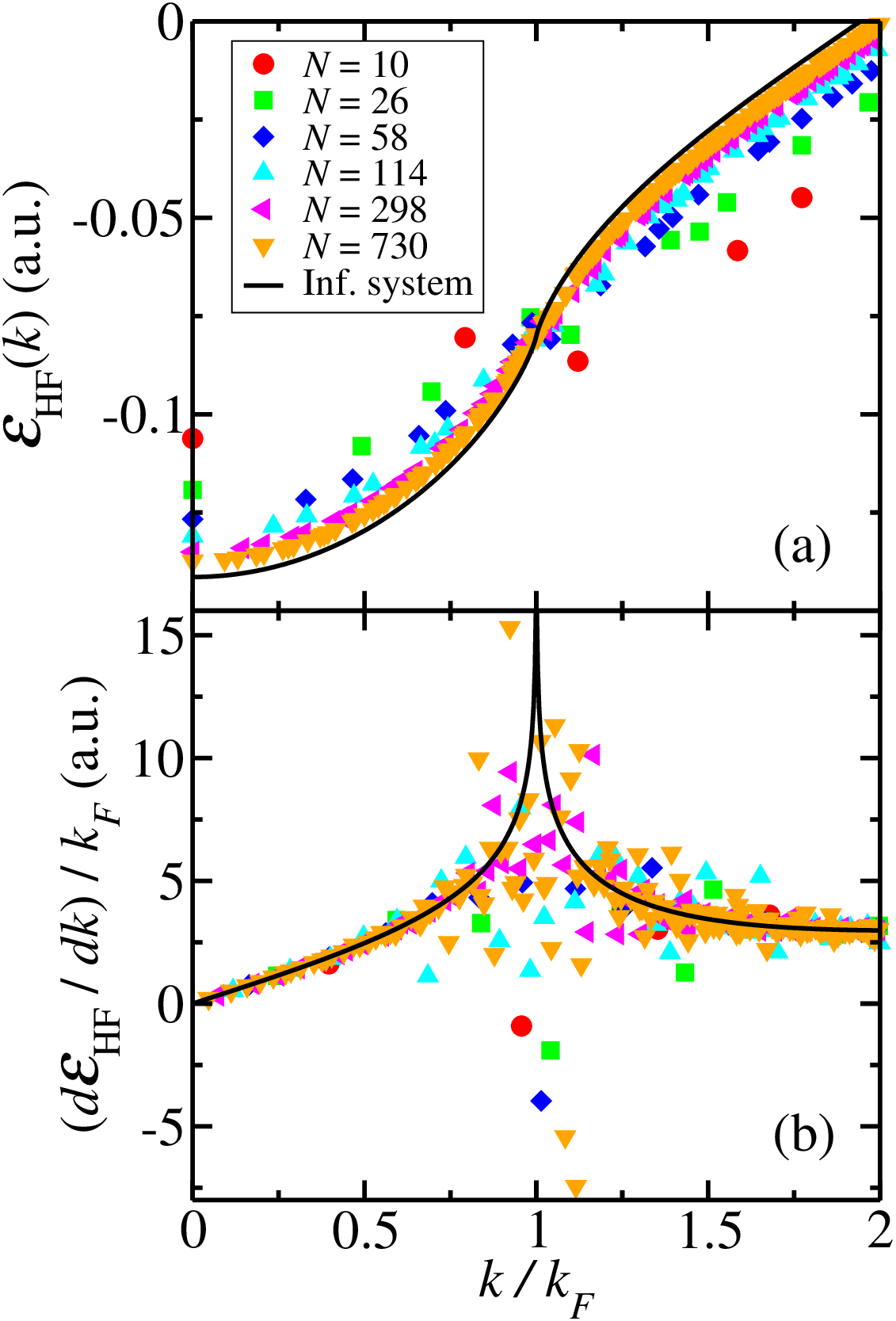}
\caption{(Color online) HF energy bands (a) and their derivatives (b)
for $N$-electron paramagnetic 2D HEGs of density parameter $r_s=10$
a.u.
  \label{fig:para_rs10_HF_band}}
\end{center}
\end{figure}

\begin{figure}
\begin{center}
\includegraphics[scale=0.80,clip]{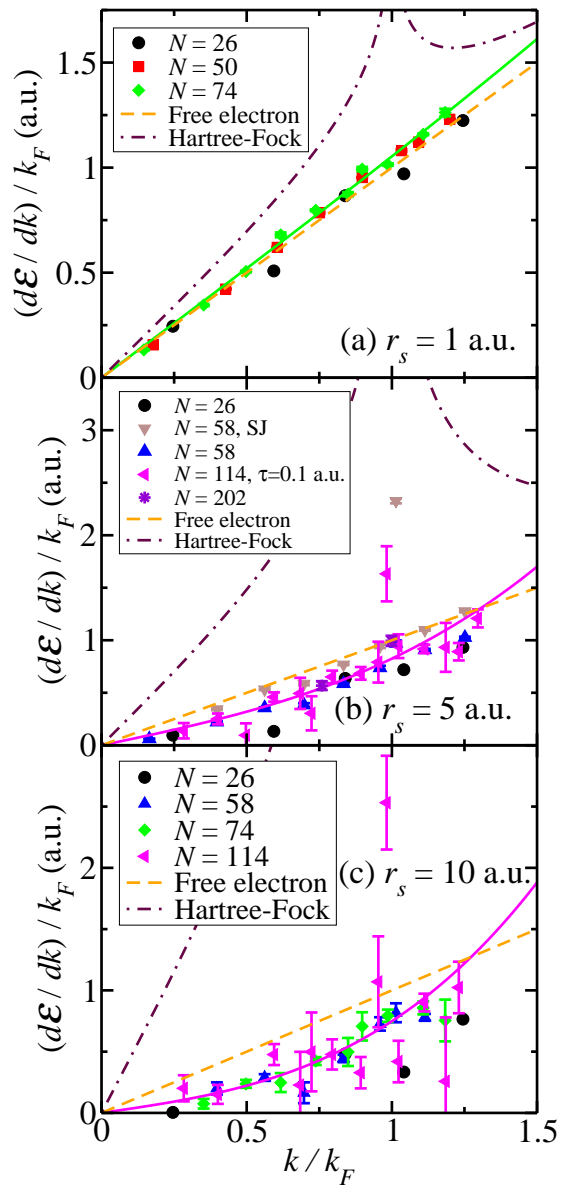}
\caption{(Color online) Derivatives of the energy bands of
paramagnetic $N$-electron 2D HEGs at $r_s=1$ a.u.\ (top), 5 a.u.\
(middle), and 10 a.u.\ (bottom). The curve labeled ``SJ'' used a
Slater-Jastrow trial wave function; the others used a
Slater-Jastrow-backflow trial wave function.  Except where indicated
otherwise, DMC time steps $\tau$ of 0.01, 0.2, and 0.4 a.u.\ were used
at $r_s=1$, 5, and 10 a.u.  The central difference approximation was
used to evaluate the numerical derivatives. The solid lines show the
derivatives of quartic fits to the DMC energy bands for $N=74$, 114,
and 114 electrons at $r_s=1$, 5, and 10 a.u., respectively.
\label{fig:para_dbands}}
\end{center}
\end{figure}

\begin{figure}
\begin{center}
\includegraphics[scale=0.80,clip]{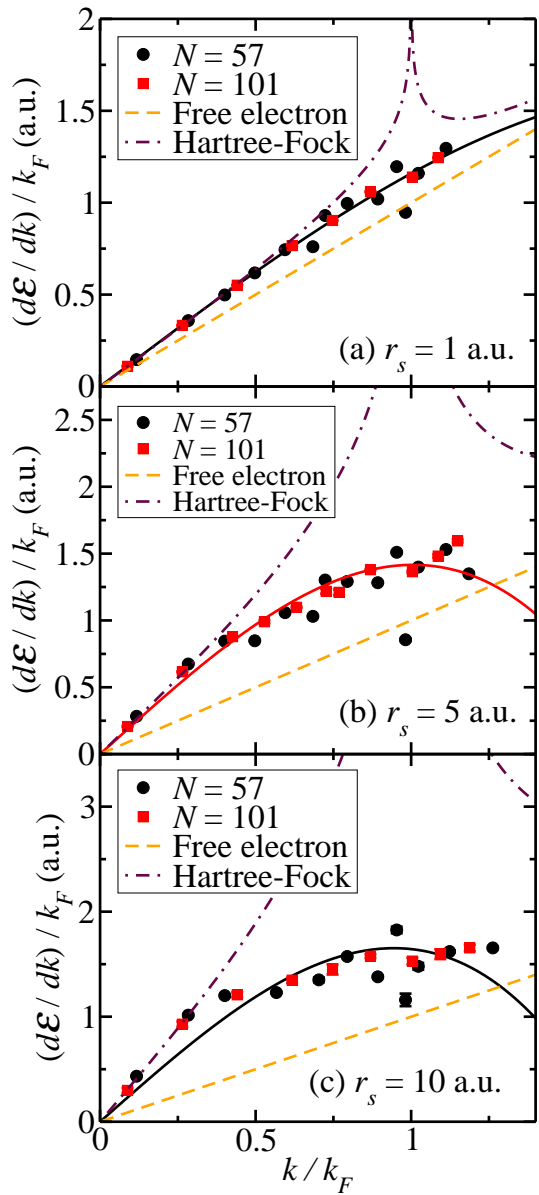}
\caption{(Color online) As Fig.\ \ref{fig:para_dbands}, but for
ferromagnetic HEGs. The solid lines show the derivatives of quartic
fits to the DMC energy bands for $N=57$, 101, and 57 electrons at
$r_s=1$, 5, and 10 a.u., respectively.
  \label{fig:ferro_dbands}}
\end{center}
\end{figure}

A possible explanation for this difference is that we obtained our
effective masses by fitting a band to a wide range of $k$ values
instead of just considering the behavior in the vicinity of the Fermi
surface.  As can be seen in Fig.\ \ref{fig:para_rs10_HF_band}, the HF
bands of finite systems as well as infinite ones exhibit pathological
behavior near the Fermi surface. The derivatives of the bands become
large and fluctuate wildly.  The numerical derivatives of the DMC
bands are plotted in Figs.\ \ref{fig:para_dbands} and
\ref{fig:ferro_dbands}.  It can be seen that the DMC bands exhibit
residual Hartree-Fock-like pathological behavior at the Fermi surface,
because not all the correlation energy is retrieved.  The pathological
behavior is more pronounced in paramagnetic HEGs and at low densities.
Our procedure of fitting the band over a wide range of $k$ enables us
to avoid this pathological behavior.  The finite-size effects
considered by Holzmann \textit{et al.}\ only affect the energy band in
the vicinity of the Fermi surface, suggesting that it is also possible
to avoid these finite-size effects by fitting the band over a wide
range of $k$.

It may seem counterintuitive to use DMC data for excitations far from
the Fermi surface in order to determine the effective mass, which is a
parameter in a theory describing excitations in the vicinity of the
Fermi surface.  However, we reiterate that it is a premise of Fermi
liquid theory that the quasiparticle band coincides with the energy
band defined by differences in total-energy eigenvalues in the
vicinity of the Fermi surface, so that the derivative of the
quasiparticle band at the Fermi surface is equal to the derivative of
the electronic energy band.  It is possible in principle that the
gradient of the energy band may change sharply in the vicinity of the
Fermi surface, but there seems to be no reason to suppose this to be
the case and our results do not provide any evidence for this sort of
behavior: the energy bands shown in Figs.\ \ref{fig:para_bands} and
\ref{fig:ferro_bands} look well-behaved.  Nevertheless, it is
comforting to observe that our effective-mass results are insensitive
to a reduction in the range of $k$-values used to perform the fit.
This result is obvious from looking at Figs.\ \ref{fig:para_bands} and
\ref{fig:ferro_bands}, and it also follows from a quantitative study,
as shown in Tables \ref{table:para_fit_range} and
\ref{table:ferro_fit_range}.  For both the paramagnetic and
ferromagnetic HEGs, discarding data far from $k_F$ has little effect
on the calculated effective mass, until the remaining data points are
all sufficiently close to $k_F$ that either the pathological behavior
of the DMC energy band in this region starts to dominate or there are
insufficient data to perform an accurate fit.

\begin{table}
\begin{center}
\begin{tabular}{ccr@{.}l} \hline \hline

Range & No.\ pts in fit & \multicolumn{2}{c}{$m^\ast$ (a.u.)} \\

\hline

$0.0~ \leq k/k_F \leq 1.33$ & 17 & ~$1$&$21(2)$ \\

$0.23 \leq k/k_F \leq 1.26$ & 15 &  $1$&$18(2)$ \\

$0.33 \leq k/k_F \leq 1.20$ & 13 &  $1$&$14(2)$ \\

$0.47 \leq k/k_F \leq 1.17$ & 11 &  $1$&$12(3)$ \\

\hline \hline
\end{tabular}
\caption{Effective mass $m^\ast$ versus range of $k$ values used to
fit the energy band for a 114-electron paramagnetic HEG at $r_s=5$
a.u. \label{table:para_fit_range}}
\end{center}
\end{table}

\begin{table}
\begin{center}
\begin{tabular}{ccr@{.}l} \hline \hline

Range & No.\ pts in fit & \multicolumn{2}{c}{$m^\ast$ (a.u.)} \\

\hline

$0.0~ \leq k/k_F \leq 1.18$ & 12 & ~$0$&$706(9)$ \\

$0.18 \leq k/k_F \leq 1.12$ & 10 &  $0$&$72(1)$  \\

$0.35 \leq k/k_F \leq 1.06$ & ~8 &  $0$&$722(9)$ \\

$0.5~ \leq k/k_F \leq 0.95$ & ~6 &  $0$&$686(8)$ \\

\hline \hline
\end{tabular}
\caption{Effective mass $m^\ast$ versus range of $k$ values used to
fit the energy band for a 101-electron ferromagnetic HEG at $r_s=5$
a.u. \label{table:ferro_fit_range}}
\end{center}
\end{table}

\begin{figure}
\begin{center}
\includegraphics[scale=0.29,clip]{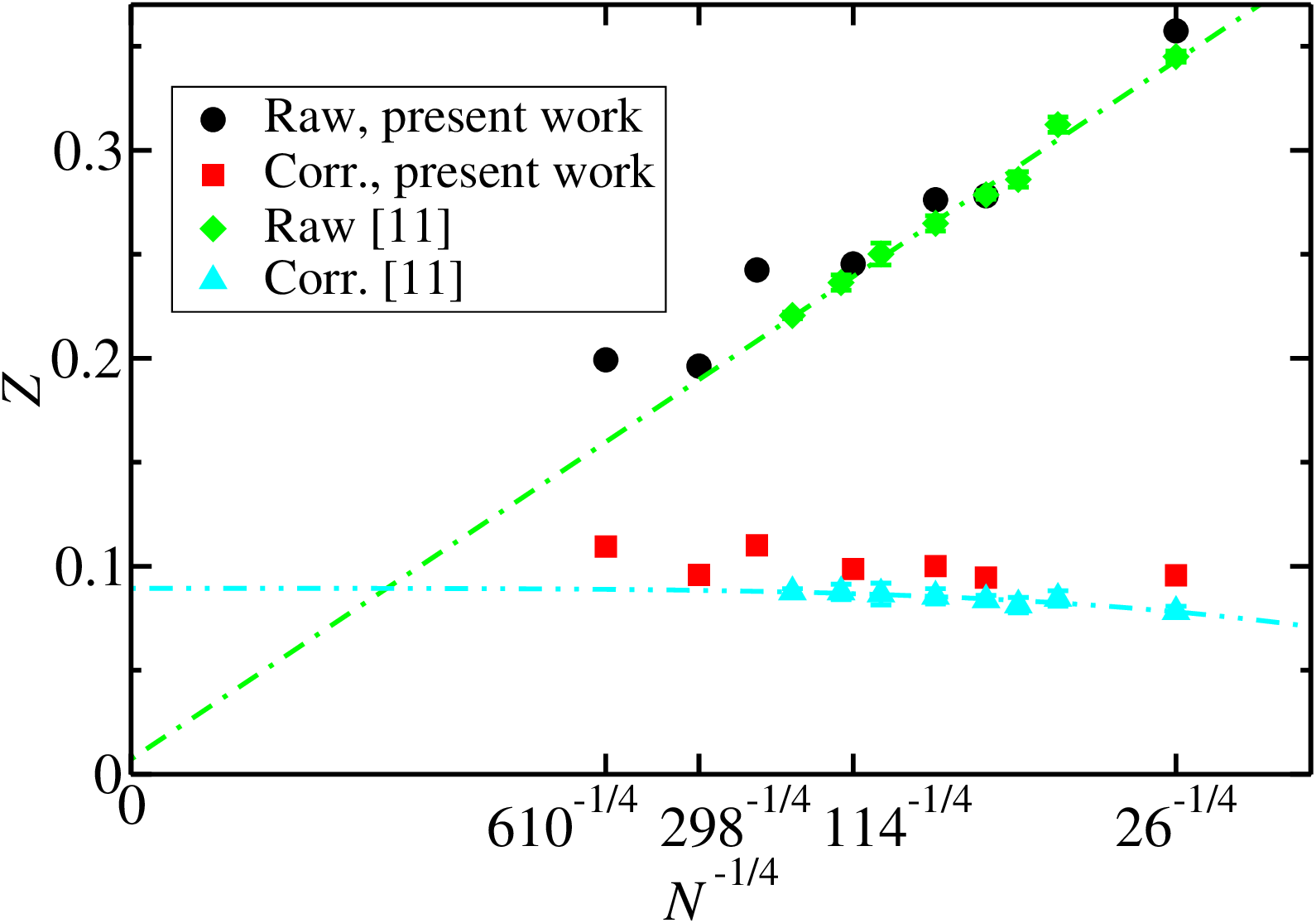}
\caption{(Color online) VMC renormalization factor $Z$ against number
of electrons $N$ for paramagnetic 2D HEGs at $r_s=10$
a.u. \label{fig:Z_v_N}}
\end{center}
\end{figure}

\begin{figure}
\begin{center}
\includegraphics[scale=0.33,clip]{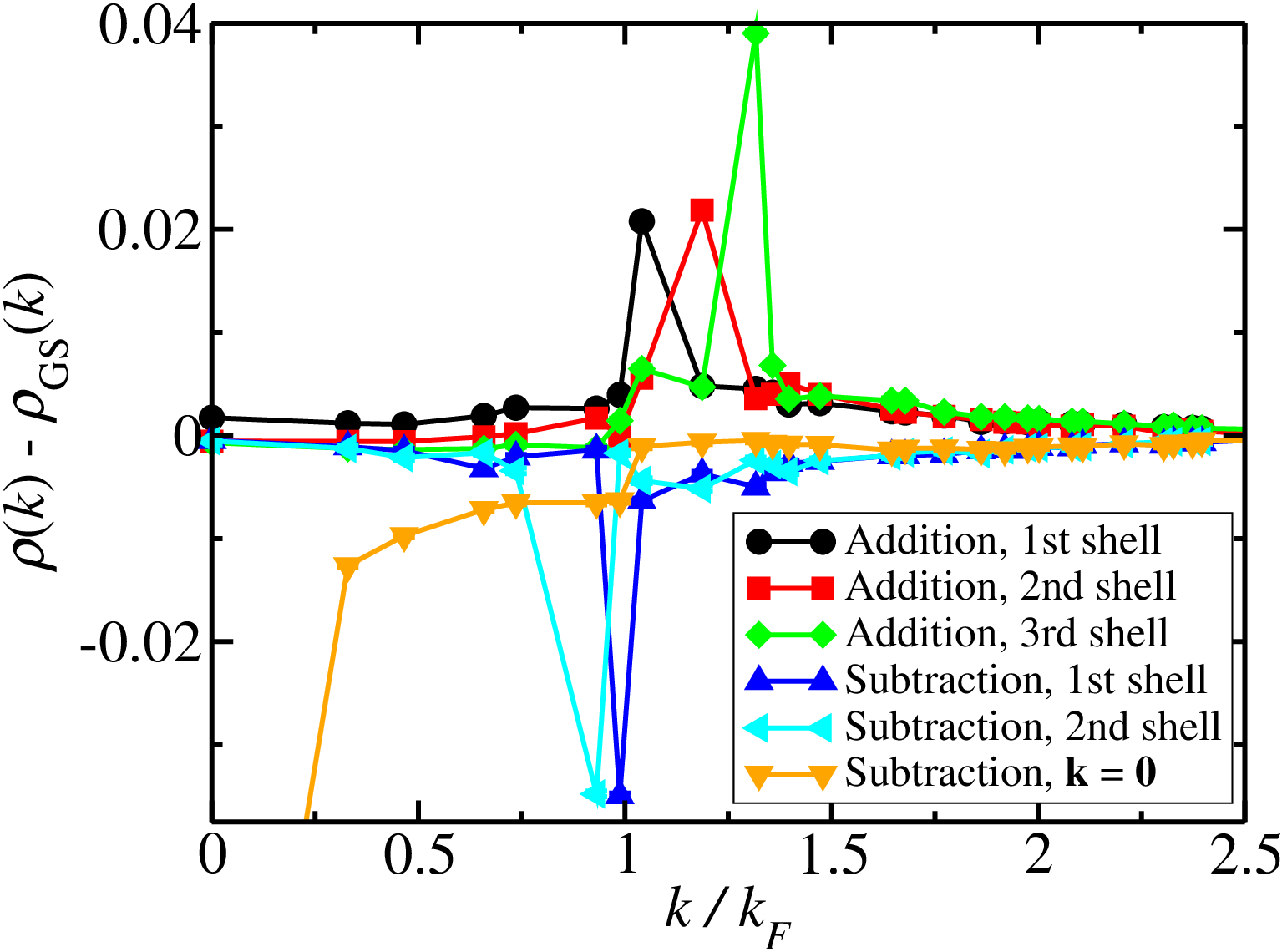}
\caption{(Color online) VMC momentum density $\rho(k)$ relative to the
ground-state momentum density $\rho_{\rm GS}(k)$ for different
excitations to a paramagnetic 58-electron HEG at $r_s=10$ a.u. The
momentum densities are averaged over reciprocal-lattice vectors of the
same length, so the height of the spike at the $|{\bf k}|$ at which
the electron is added looks smaller when there are many
reciprocal-lattice vectors with the same length as $|{\bf
k}|$. \label{fig:rel_momdist}}
\end{center}
\end{figure}

Another possible reason for not reproducing the finite-size errors in
the effective mass predicted by Holzmann \textit{et al.}\ might simply
be that some of their assumptions are invalid.  We have reproduced
their $O(N^{-1/4})$ scaling of the finite-size error in the
renormalization factor (the discontinuity $Z$ in the momentum density
at the Fermi edge) in VMC calculations, as shown in Fig.\
\ref{fig:Z_v_N}.  However, when an electron is added, there are two
contributions to the momentum density: a peak of weight $Z$ at the
momentum at which the electron is added and a smeared-out background
of weight $1-Z$.\cite{giuliani} Together, these two contributions to
the change in the momentum density are responsible for the change in
the kinetic energy when the electron is added, i.e., for the kinetic
contribution to the energy band.  Examples of the changes in the VMC
momentum density that result from adding or removing electrons from
different ${\bf k}$ are shown in Fig.\ \ref{fig:rel_momdist}.  It can
be seen that the smeared-out background depends on the ${\bf k}$ at
which the electron is added, its average tending to increase with $k$.
The finite-size error in the weight $Z$ of the peak is equal and
opposite to the finite-size error in the weight $1-Z$ of the
background.  The finite-size error in the derivative of the energy
band due to the background therfore tends to cancel the finite-size
error due to the peak.  So it is not clear that the $O(N^{-1/4})$
finite-size error in the renormalization factor should result in an
$O(N^{-1/4})$ error in the gradient of the energy band and hence
effective mass.

\section{Conclusions \label{sec:conclusions}}

In summary, we have used DMC to calculate the energy band of the
interacting 2D HEG, and hence we have obtained the quasiparticle
effective mass.  Our ferromagnetic and paramagnetic effective masses
are in agreement with the experimental results of Padmanabhan
\textit{et al.}\cite{padmanabhan}\ and Tan \textit{et
al.},\cite{tan2005} respectively.  In particular, our data confirm
that the effective mass of the paramagnetic HEG increases when the
density is lowered, while the effective mass of the ferromagnetic HEG
decreases.

\begin{acknowledgments} We acknowledge financial support from the
Leverhulme Trust, Jesus College, Cambridge, and the UK Engineering and
Physical Sciences Research Council.  Computing resources were provided
by the Cambridge HPCS and HPCx.  We are grateful to W.\ M.\ C.\
Foulkes for many useful discussions.  We thank M.\ Padmanabhan and
Y.-W.\ Tan for sending us their experimental effective-mass data and
R.\ Asgari for sending us his $GW$ effective-mass data.
\end{acknowledgments}


\begin{thebibliography}{99}
\bibitem{footnote_k_F} The Fermi wave vector is $k_F=\sqrt{2}/r_s$ for
the 2D paramagnetic fluid and $2/r_s$ for the 2D ferromagnetic fluid.
\end{thebibliography}
\end{document}